\documentclass[twocolumn,prl, showpacs,showpacs,aps]{revtex4}
\usepackage{graphicx}
\usepackage{dcolumn}
\usepackage{bm}
\usepackage{amsmath}
\usepackage{amssymb}
\usepackage{epsf}
\newcommand{\be}{\begin{equation}}
\newcommand{\ee}{\end{equation}}
\newcommand{\bea}{\begin{eqnarray}}
\newcommand{\eea}{\end{eqnarray}}

\newcommand{\LD}{$\Lambda$-doublet}

\begin{document}
\title{Magneto-electrostatic trapping of ground state OH molecules}
\author{Brian C. Sawyer}
 \email{sawyerbc@colorado.edu}
\author{Benjamin L. Lev}
\author{Eric R. Hudson}\altaffiliation{Present address: Department of Physics, Yale University, New Haven, CT 06520, USA}
\author{Benjamin K. Stuhl}
\author{Manuel Lara}
\author{John L. Bohn}
\author{Jun Ye}
\affiliation{JILA, National Institute of Standards and Technology
and the University of Colorado \\ Department of Physics, University
of Colorado, Boulder, Colorado 80309-0440, USA}
\date{\today}
\begin{abstract}
We report the magnetic confinement of neutral, ground state hydroxyl
radicals (OH) at a density of $\sim$3$\times10^{3}$ cm$^{-3}$ and
temperature of $\sim$30 mK. An adjustable electric field of
sufficient magnitude to polarize the OH is superimposed
on the trap in either a quadrupole or homogenous field geometry. The
OH is confined by an overall potential established via molecular
state mixing induced by the combined electric and magnetic fields
acting on the molecule's electric dipole and magnetic dipole moments,
respectively. An effective molecular Hamiltonian including Stark and
Zeeman terms has been constructed to describe single molecule
dynamics inside the trap.  Monte Carlo simulation using this
Hamiltonian accurately models the observed trap dynamics in various
trap configurations. Confinement of cold polar molecules in a
magnetic trap, leaving large, adjustable electric fields for
control, is an important step towards the study of low energy
dipole-dipole collisions.

\end{abstract}
 \maketitle

Cold and ultracold molecules are currently subject to intense
research efforts as they promise to lead major advances in precision
measurement, quantum control, and cold chemistry~\cite{Doyle04}.
Much of this interest has focused on molecules possessing permanent
electric dipole moments. The long range, anisotropic dipole-dipole interaction leads to novel collision~\cite{BohnFieldLinked,GerardOHXe} and reaction~\cite{HudsonH2CO}
dynamics, and may serve as the interaction between molecular
qubits~\cite{LevPRA} in architectures for quantum information
processing~\cite{DemilleQC}. Quantum phase transitions can be
explored with polar molecules~\cite{Zoller06}. Furthermore, polar
molecules may possess extremely large internal electric fields (on the
order of GV/cm), yielding dramatic sensitivity enhancement to
searches for an electron electric dipole
moment~\cite{HindsEDM,DemilleEDM}. Cold polar molecules have been
produced via photoassociation of ultracold atomic
species~\cite{JulienneReview,DeMille05}, Stark
deceleration~\cite{Meijer99,Bochinski03}, and buffer gas
cooling~\cite{Doyle98}.
\begin{figure}[ht]
\scalebox{.45}[.45]{\includegraphics{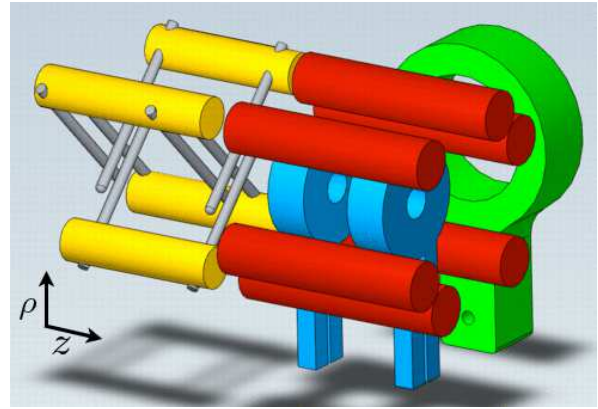}}
\caption{\label{fig1}(color online) Illustration of the
magneto-electrostatic trap (MET) assembly.  The terminus of the
Stark decelerator (yellow, gray electrodes), shown to the left,
couples state-selected cold molecules into the magnetic quadrupole
(blue). The double-electrode structures within the electric
quadrupole (red) allow for application of uniform electric fields
within the magnetic trap. The ring (green) supports a 1 inch lens
for collection of laser-induced fluorescence.}
\end{figure}

Polar molecules are readily trapped in inhomogeneous electric fields
~\cite{MeijerETrap,RempeETrap}, however theory suggests that at high
phase-space densities such traps invariably suffer from large
inelastic collision losses~\cite{Bohn02}.  One possible solution is
to trap strong-field seeking states via time-dependent electric
traps ~\cite{Veldhoven05}.  In contrast, we are interested in magnetically trapping these neutral polar molecules so as to have the freedom to apply
external electric fields to control collision dynamics inside the
trap. Highly vibrationally excited, triplet KRb molecules produced via
photoassociation have been magnetically trapped at a temperature of
$\sim$$300$ $\mu$K and density of $10^{4}$
cm$^{-3}$~\cite{EylerTrap}. Buffer gas cooling has been used to
load a magnetic trap with ground-state CaH and NH
molecules at temperatures of hundreds of mK and densities of
$\sim$$10^{8}$ cm$^{-3}$, but in the presence of a cold He buffer gas~\cite{Doyle98,Doyle07}. We report here the first observation
of magnetically trapped hydroxyl radical (OH) molecules and the
lowest temperature (30 mK) yet achieved for a magnetically-trapped
polar molecule in its rovibronic ground state.  Moreover, we perform
the first study of trap dynamics based on a single molecule
possessing large magnetic and electric dipole moments in B- and
E-fields that are inhomogeneous and anisotropic.
Interesting polar molecule dynamics have been predicted at large field
strengths~\cite{Tscherbul06}. Accurate correspondence between
experimental data and simulations is achieved only by accounting for
the molecular state mixing induced by the crossed fields. Large
magnetic (B) fields may serve to suppress inelastic collision
losses~\cite{BohnMag} while at the same time sympathetic cooling of
molecules via co-trapped ultracold atoms may become
feasible~\cite{Lara07}. In addition, the magnetic quadrupole trap
described here permits the application of arbitrary external
electric (E) fields to a large class of polar molecules---a necessary step towards observing cold molecule
dipole-dipole collisions~\cite{Krems06}.

To magnetically trap OH, we begin with a cold beam of ground state
($^{2}\Pi_{3/2}$) OH  produced via a pulsed electric discharge
through $2\%$ H$_{2}$O vapor~\cite{Lewandowski04} seeded in 1.5 bar
of Kr. A piezoelectric transducer valve~\cite{PZT} operating at 10
Hz provides a supersonic expansion of the OH/Kr mixture through a 1
mm nozzle. This results in a 490 m/s molecular packet with a 15\%
longitudinal ($\hat{z}$) velocity spread. The transverse
($\hat{\rho}$) velocity spread of the beam is limited to 2\% by the
3 mm diameter molecular skimmer.  Compared with a Xe expansion, the
higher-velocity Kr expansion leads to less transverse beam loss
within the current decelerator design. Thus, despite the fact that
the Kr beam requires a higher phase angle for deceleration (smaller
decelerator acceptance) it results in more decelerated molecules.
After passing through the skimmer, the OH packet is spatially
matched to the decelerator entrance via an electrostatic hexapole.
Details regarding the OH Stark decelerator design and operating
principle may be found in previous work~\cite{Bochinski04,Hudson04}.
We utilize the 142 stages of our Stark decelerator to slow a 130 mK
packet (in the co-moving frame) of OH to 20 m/s at a phase angle
$\phi_{0}=47.45^{\circ}$ for coupling into the magneto-electrostatic
trap (MET).  Decelerated OH packets are detected via laser-induced
fluorescence (LIF) upon entering the MET, whose center lies 2.7 cm
from the final pair of decelerator rods. The 282 nm pulsed laser
beam is oriented along the mutual longitudinal axis ($\hat{z}$) of
the decelerator and MET, and the resulting 313 nm LIF is collected
in a solid angle of 0.016 sr.
\begin{figure}[Potentials]
\scalebox{.42}[.42]{\includegraphics{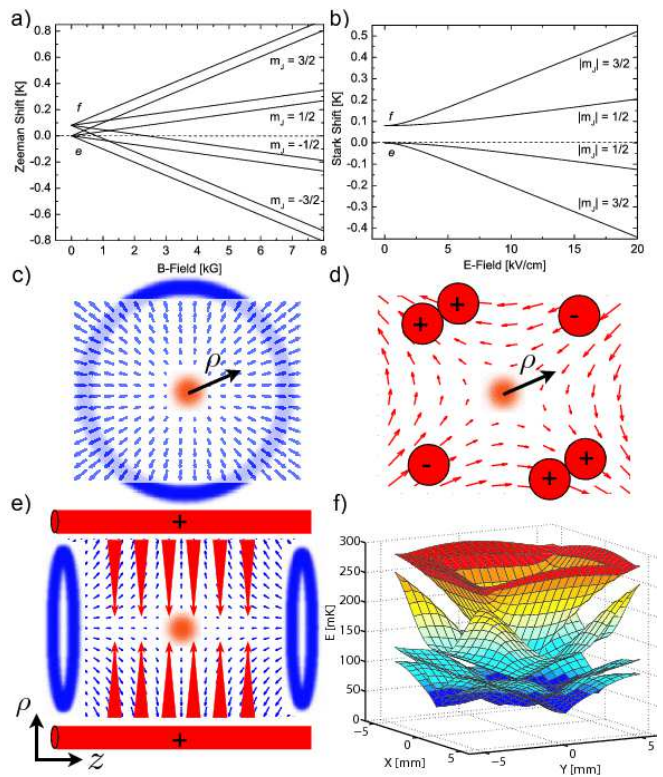}}
\caption{\label{Potentials}(color online) (a) Zeeman and (b) Stark
effects of the ground state structure of OH. (c) Magnetic and (d)
electric quadruple fields viewed from $\hat{z}$. (e) Side view of
the MET configuration with quadrupole E and B fields. (f) Transverse
adiabatic potential surfaces for various components of the OH ground
state at the longitudinal MET center. The top surface depicts the
decelerated/trapped $|J=\frac{3}{2},m_{J}=\frac{3}{2}\rangle$
state.}
\end{figure}

The MET is illustrated in Fig.~\ref{fig1}, and consists of two
copper coils operated in an anti-Helmholtz configuration surrounded
by an electric quadrupole. The center-to-center magnetic coil
spacing is 1.5 cm, providing a field gradient of 6700 G/cm by
passing 1500 A through the 4 turns of each water-cooled coil. A
longitudinal trap depth of 400 mK is achieved, while the electrodes
surrounding the coils allow the application of an E-field to the
trapped sample. The double- and single-rod electrodes that comprise
the electric quadrupole allow the application of two distinct field
configurations. A uniform bias electric field is created by
grounding the single-rods while oppositely charging the opposing
double-rods to $\pm3$ kV. Alternatively, we generate a quadrupole E-field by
charging neighboring rods with opposite polarities of $\pm9$ kV.  Shielding by the MET coils reduces the interior E-field by a factor of $\sim$5.  Figure~\ref{Potentials} depicts the quadrupole electric field
superimposed on the magnetic quadrupole field.  OH has both an
appreciable magnetic and electric dipole, and we can easily alter
the geometry of the trapping potential by applying electric fields
of variable magnitude and direction to the magnetically trapped
sample.

The OH ground state is best described by Hund's case ($a$), in which
the spin ($|\Sigma|=\frac{1}{2}$) and orbital ($|\Lambda|=1$)
angular momenta are strongly coupled to the molecular axis,
yielding the total angular momentum projection
$|\Omega|=\frac{3}{2}$. The $|\Omega|=\frac{1}{2}$ state lies 126
cm$^{-1}$ above this ground state. The nuclear spin of the hydrogen
atom ($I=\frac{1}{2}$) leads to a
hyperfine splitting of the ground state into $F=2,1$ components. In
the presence of large external fields (E$>$1 kV/cm or B$>$100 G),
the electron angular momentum and nuclear spin
decouple~\cite{LevPRA}, and one considers the total angular momentum
\textbf{J}, which includes nuclear and electronic orbital angular
momentum as well as electron spin. The projections $m_{J}$ of
\textbf{J} are defined along the axis of the applied field. The
ground state is further split into two opposite-parity states
($f$/$e$) by the coupling of electronic orbital angular momentum to
nuclear rotation. This small $\Lambda$-doublet splitting of 1.67 GHz, along with the 1.67 D electric dipole, is
responsible for the rather large Stark shift experienced by OH
molecules~\cite{Hudson06}. Fig.~\ref{Potentials}(a) and (b) show the
Zeeman and Stark effects of the OH ground-state respectively.
External electric fields can thus dramatically alter the potential
confining the magnetically trapped molecules.
Fig.~\ref{Potentials}(c-e) illustrate the field geometries
within the MET.  In the plane perpendicular to $\hat{z}$, Fig.~\ref{Potentials}(c) shows the radial B-field
characteristic of a magnetic quadrupole, while
Fig.~\ref{Potentials}(d) depicts the electric quadrupole field
introduced by the electrode geometry. Fig.~\ref{Potentials}(e)
illustrates both B (blue) and E (red) fields along $\hat{z}$.

The top potential surface of Fig.~\ref{Potentials}(f) represents the
energy shift of the decelerated
$|J=\frac{3}{2},m_{J}=\frac{3}{2}\rangle$ state (E-fields couple $e$
and $f$, therefore they are no longer good quantum numbers) within
the combined quadrupole E- and B-fields present in the MET. The
transverse profiles of the four trapping adiabatic potentials are
depicted at the longitudinal trap center, where
$\overrightarrow{B}\cdot\hat{z}=0$. The remaining four surfaces (not
shown) are simply inverted relative to those displayed, and are
therefore anti-trapping. The three-dimensional MET potentials are
calculated by diagonalizing an effective Hamiltonian that includes
both Stark and Zeeman terms. Coupling between the ground
$|J=\frac{3}{2},\Omega=\frac{3}{2}\rangle$, and excited
$|\frac{5}{2},\frac{3}{2}\rangle$,
$|\frac{1}{2},\frac{1}{2}\rangle$, $|\frac{3}{2},\frac{1}{2}\rangle$
states is included, for a total of 64 hyperfine components. The
$|J=\frac{3}{2},\Omega=\frac{3}{2}\rangle$,
$|\frac{3}{2},\frac{1}{2}\rangle$ coupling is the most critical,
increasing the trap depth of the top surface of
Fig.~\ref{Potentials}(f) by 11\%. Addition of the remaining states
modifies the potential by $<$0.1\%. The square shape of the top
surface is a result of the varying angle between E and B in the MET.
The potentials from each field directly sum where the fields are
collinear, but the trap shallows from this maximum value as the
angle between E and B increases.

The Stark decelerator slows only those OH molecules in the
electrically weak-field seeking (EWFS) component of the ground
state, \textit{i.e.}, the upper \LD. Of these EWFS molecules, only
those with components $J=\frac{3}{2}$, $m_{J}=\pm\frac{3}{2}$ are
synchronously slowed to 20 m/s.  Because the positive $m_{J}$ state
is magnetically weak-field seeking, only 50\% of the decelerated
molecular packet is trappable by the MET.  The molecules experience
a B-field $>$100 G at the last deceleration stage emanating from the
back coil, preserving the quantization of the decelerated state.

To trap the incoming OH packet, the coil closer to the decelerator
(front coil) is grounded while the coil further from the decelerator
(back coil) operates at 2000 A for the duration of the 3.7 ms
deceleration sequence. The effective magnetic dipole of OH in its
$J=m_{J}=\frac{3}{2}$ state is $1.2$ $\mu_{B}\simeq0.81$ K/T, where
$\mu_{B}$ is the Bohr magneton, allowing the back coil to stop
molecules with velocity $\leq$ 22 m/s. The front coil is turned on
after the 20 m/s molecules are stopped by the back coil, which
occurs 2.65 ms after exiting the final deceleration stage. The trap
switching is synchronous to the 20 m/s molecules since OH possessing
larger longitudinal velocities escape the quadrupole field.  For
technical reasons, the two magnet coils are connected in series,
causing the current in both coils to decrease from 2000 A to 1500 A
over $\sim$800 $\mu$s once the front coil is switched into the
circuit. This problem will be addressed with a current servo.
\begin{figure}[ToFData]
\scalebox{.4}[.4]{\includegraphics{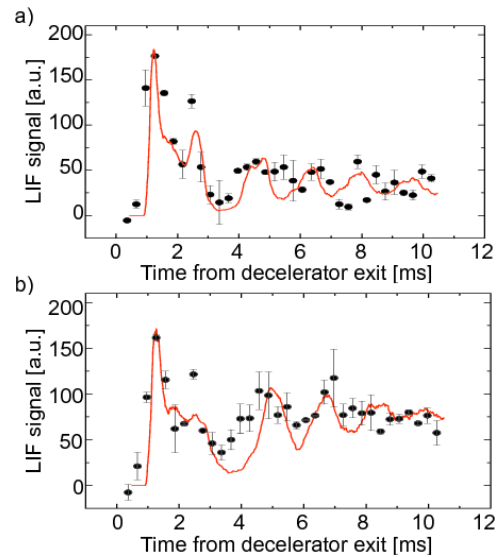}}
\caption{\label{TOF}(color online) Time-of-flight data and Monte
Carlo simulation results (solid red line) for two different MET
configurations.  The magnetic field switches from the stopping
configuration to quadrupole trapping at $t=2.65$ ms.  (a) Magnetic
trap only.  (b) Stopping and trapping in the presence of combined
electric and magnetic quadrupole fields.  Note larger steady-state
trap population.}
\end{figure}

Figure~\ref{TOF} displays typical time-of-flight data resulting from
the trapping sequence along with results from Monte Carlo simulation
of the trap dynamics. The large initial peak is the decelerated OH
packet traversing the detection region at the center of the trap.
All subsequent peaks are the result of oscillation in the
$\hat{\rho}$ dimension, with a frequency of $\sim$650 Hz.
Fluorescence is gathered in $\hat{\rho}$, and oscillations in
$\hat{z}$ are not detectable since our detection region spans the
entire visible trap length. The magnetic quadrupole trap is unable
to confine all the molecules stopped by the back coil alone due to
the 25\% current drop, as is clear from the difference in signal
height at $\sim$2.5 ms versus the steady-state level at, e.g., 10
ms. Note the larger steady-state population of Fig.~\ref{TOF}(b)
from the addition of a confining transverse electric quadrupole.
Also visible is the shorter coherence time compared to
Fig.~\ref{TOF}(a), which results from trap distortion visible in
Fig.~\ref{Potentials}(f). An estimate of the density of trapped OH,
as measured by LIF, gives $\sim3\times$10$^{3}$ cm$^{-3}$.  By
comparison, the density of a packet ``bunched" at a phase angle of
$\phi_{0}=0^{\circ}$ is $10^{7}$ cm$^{-3}$, while the 20 m/s packet
is $10^{4}$ cm$^{-3}$.  Monte Carlo analysis of the velocity
distribution of the trapped OH yields temperatures of $\sim$30 mK.
With our improved knowledge of the trap dynamics due to accurate
Monte Carlo simulations, we expect to be able to trap 2-3 times more
OH through modification of deceleration phase angle and trap
switching times for optimal coupling of the slowed beam into the
MET.

Higher decelerator phase angles should, in principle, produce a slower
OH packet. However, below 20 m/s the OH population drops sharply and no further deceleration is observed. We attribute this effect to a
reduction in the effective aperture seen by slow molecules as they
traverse the final deceleration stage. The OH closest to the final
decelerator rods see a larger longitudinal potential than those
on-axis, and at sufficient phase angles, are stopped or reflected
and hence cannot be observed in the MET region. Numerical models
elucidate this effect, which may be mitigated by new decelerator
designs~\cite{Sawyer07}.
\begin{figure}[Lifetime]
\scalebox{.7}[.7]{\includegraphics{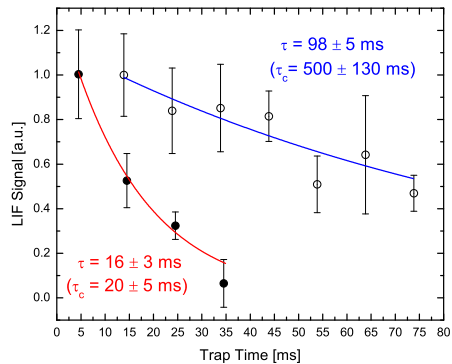}}
\caption{\label{Lifetime}(color online) Measured ($\tau$) and
de-convolved collisional ($\tau_{c}$) lifetimes of magnetically
trapped OH at a background pressure of $1\times10^{-6}$ Torr
($\bullet$, red line) and $4\times10^{-8}$ Torr ($\circ$, blue
line).}
\end{figure}

The 10 Hz repetition rate of our Stark decelerator currently limits
our trap interrogation time to 100 ms. Maintaining continuous
currents of 1500 A presents technical challenges due to heating in
both the switching transistors and cabling.  To mitigate this, we
operate the switching MOSFETs on a water-cooled copper plate and
also run chilled water through the MET coils. Water-cooled cables
are also necessary for such high-power operation.
Figure~\ref{Lifetime} shows electric field-free OH trap lifetimes
measured in two different background pressure regimes:
$1\times10^{-6}$ and $4\times10^{-8}$ Torr, both dominated by N$_2$
at 295 K.  Non-ideal power supply control causes the trap current to
decrease from 1500 A to 1160 A at a rate of 2000 A/s.  This lowers
the trap depth, resulting in a loss of $12\pm2$ s$^{-1}$.  After
accounting for this loss mechanism, we find  background
collision-limited lifetimes of $20\pm5$ ms and $500\pm130$ ms for
the two pressure regimes, respectively.  These lifetimes determine
an OH-N$_2$ collision cross section of $500\pm100$ \AA$^{2}$.  From
this clear dependence of trap lifetime on background gas pressure,
we envision measuring dipole-dipole collision cross sections by
using polar molecules as background and polarizing them with the
E-field of the MET.

We have magnetically confined ground state hydroxyl radicals in a
MET that allows for the application of a variable E-field to the
trapped polar molecules. Diagonalization of an effective Hamiltonian
modelling the energy shift of OH in combined E- and B-fields was
necessary for accurate Monte Carlo simulation of the trap dynamics.
Future experiments could quantify temperature dependent
non-adiabatic transition rates between the different surfaces
depicted in Fig.~\ref{Potentials}(f) akin to Majorana transitions
observed in atomic magnetic quadrupole traps. Furthermore, higher
densities within the MET will facilitate the study of cold OH-OH
collisions in variable electric fields. As rather modest electric
fields are expected to modify collision cross sections between polar
molecules by as much as 10$^{3}$~\cite{Bohn02}, crossed-beam
experiments exploiting the MET described here will benefit from both
low center-of-mass collision energies and tunable E-fields in the
interaction region.

We acknowledge DOE, NIST, and NSF for support. We thank P. Beckingham, H. Green, and E. Meyer for technical assistance. B. Lev is a NRC
postdoctoral fellow.
%\bibliography{MagTrap}

\end{document}